# Optimum Design of Printable Tunable Stiffness Metamaterial for Bone Healing

Mohammad Saber Hashemi[1], Karl H. Kraus[2], Azadeh Sheidaei[1*]

[1] Aerospace Engineering Department, Iowa State University, Ames, IA 50011, United States

[2] Veterinary Clinical Sciences, Iowa State University, Ames, IA 50011, United States

## Abstract

A tunable stiffness bone rod was designed, optimized, and 3D printed to address the common shortcomings of existing bone rods in the healing of long fractured bones. The common deficiencies of existing bone fixations are high stiffness, thereby negligible flexibility in deformation for best bone growth results, and stress-shielding effect. Our novel design framework provides the surgeons with ready-for-3D-printing patient-specific designs, optimized to have desired force-displacement response with a stopping mechanism for preventing further deformation under higher than usual loads such as falling. The framework is a design optimization based on the multi-objective genetic algorithm (GA) optimization to quantify the objectives, tunning the varied stiffness while minimizing the maximum Mises stress of the model to avoid plastic and permanent deformation of the bone rod. The optimum design computational framework of tunable stiffness material presented in this paper is not specific for a tibia bone rod. It can be used for any application where bilinear stiffness is desirable.

## 1. Introduction

Fractures of the long bones such as femur, tibia, fibula, humerus, radius, and ulna can be fixed using a bone plate or intramedullary rod (figure 1). The bone plate is attached to the outer surface of the bone while the bone rod is inserted into the hollow canal of the long bone. Upon applying axial load on the bone, an off-axial force is exerted on the bone-plate, and the whole assembly is rotated. This rotation causes the non-uniform formation of the callus that results in the non-uniform formation of the bone [1–8]. Consequently, it makes bone fracture of the same site more likely in a future accident. Bone rods alleviate this problem partially since the bone rod, and the bone are co-axial [9,10], preventing the broken bone from rotating. Following the bone fracture, callus forms in the fracture site and gradually becomes bone. The quality of this newly formed bone highly depends on the strain level during the walking, and it should be between 8-10% for proper healing [8,11,12]. Any deformation out of this range results in poor healing of the bone.

∗Corresponding author: Azadeh Sheidaei, PhD
Aerospace Engineering Department, Iowa State University, Ames, IA 50011, United States
Tel: 515-294-2956 (O) / Fax: 515-294-3262 (O)
Email: Sheidaei@iastate.edu

Existing bone rods are circular shafts with constant stiffness, and they do not stop the deformation under excess load to prevent higher strain on the fracture site, nor do they have the initial flexibility for optimum bone growth. Therefore, a tunable stiffness bone rod is needed to control and stop the deformation. There are few examples of flexible bone rods in the literature, but in all these designs, a mechanism or extra parts, such as spring and loose/sliding screws, have been added to add flexibility [13–19]. These designs are not patient-specific, and they are unable to control the strain level.

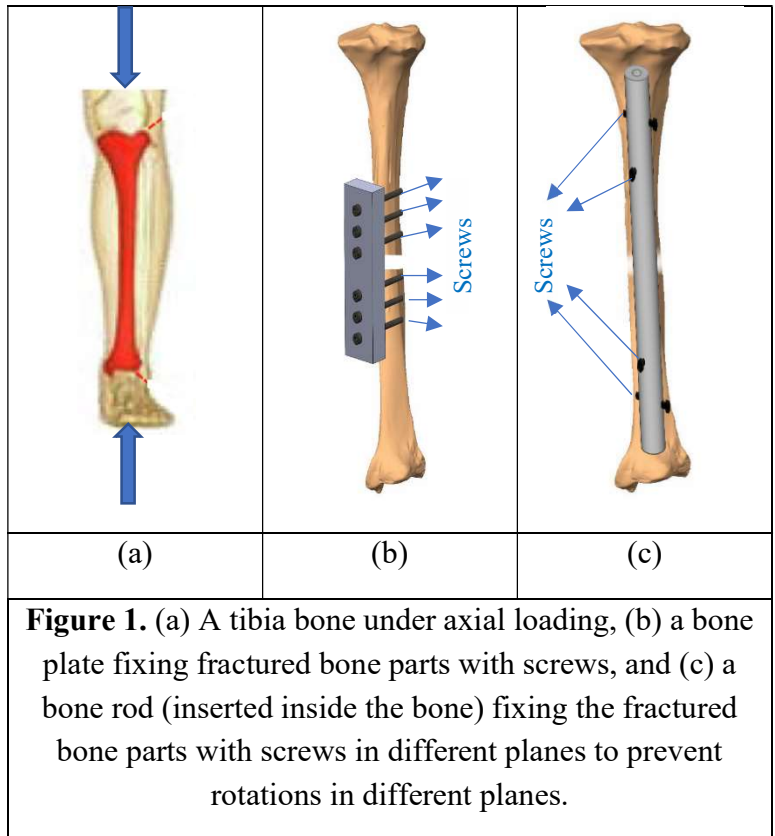

**Figure 1.** (a) A tibia bone under axial loading, (b) a bone plate fixing fractured bone parts with screws, and (c) a bone rod (inserted inside the bone) fixing the fractured bone parts with screws in different planes to prevent rotations in different planes.

In addition to the stiffness tunability of the bone fixator, its stiffness plays an essential role in the healing process. In the surgery, patients with fractures are treated using bone fixators made of stainless-steel, Cr–Co, and Ti alloys. The stiffnesses of these metals are between 110-220 GPa, which are much higher than that of human cortical bone ~20 GPa [20]. As a result, the majority of the load is carried by the fixator rather than the underlying bone. Subsequently, callus formation, ossification, and bone union at the fractured part are hindered after the implant operation, and the whole bone structure, and not just the fractured section, becomes osteoporosis. The bone mass can be decreased by 20%, and in some cases, the bone re-fracture due to stress concentration around the bone screws [21]. These phenomena are widely recognized as the "stress shielding" effect, which is the main drawback of the use of metal bone fixators [20]. A bone fixator with stiffness close to the bone or lower is needed to prevent stress-shielding and to prevent osteoporosis in patients. The objective of this study is the design and manufacture of a bone rod with tunable stiffness. In order to achieve this objective, the conceptual design of a tunable stiffness metamaterial will be presented. Then this design will be optimized by utilizing a multi-objective GA optimization [22] coupled with FE simulation methods. The final design will be a printable tunable bone rod for given patient weight and fracture gap size.

## 2. Conceptual design of a tunable stiffness metamaterial

A metamaterial is engineered to have a property that is not found in naturally occurring materials. Mechanical metamaterials attain their extreme properties from their architectures instead of their compositions [23]. Some metamaterials achieve their superior properties, such as negative Poisson's ratios [24,25], very high strength-to-weight ratios [26], and negative thermal expansion coefficients [27], by controlling the configurations of their constitutive cells. Some studies have focused on isotropic metamaterials by repeating a single cell boundlessly, which is easy to design. However, many practical applications necessitate finite and irregular volumes, as well as anisotropic properties. To overcome this challenge, some have developed data-driven design framework for making aperiodic and arbitrarily shaped metamaterials with a set of characterized deformable material layers [28] or assembling different microstructures made of a single stiff material by a 3D printer with varying compliances to achieve the desired soft deformation in different locations of a given volume space [29]. Additionally, functionally graded materials with respect to elasticity [30] and thermal expansion [31] have been developed based on aperiodic microstructural cells in specific directions. The previous studies mostly have utilized finite element (FE) analysis, sparse regularization, and constrained optimization [32]. Nevertheless, combinatorial strategy and a new approach based on the freedom and constraint topologies methodology have also been proposed to design frustration-free aperiodic spatially textured metamaterials [33] and directionally compliant metamaterials (with prescribed compliant directions and high stiffness in all other directions) [34], respectively. Compared with static 3D printed metamaterials, 4D printing opened the possibility of having an alive 3D printed structure which reacts to external stimuli such as temperature by changing its configuration with time [35]. For instance, programmable shape-changing of shape memory polymers created from digital manufacturing (3D printing) has demonstrated the ability to make devices with novel multifunctional performances [36,37]. In this study, a static elastic metamaterial with tunable bilinear stiffness in one direction and a limited volume constraint is desired. Inspired by the design of our new 2D tunable stiffness metamaterial [38], a 3D model is designed (figure 2b) by evolving the 2D schematic of an axisymmetric model shown in figure 2a.

Upon applying axial force on this model, first, it deforms with low stiffness (figure 2c), and it becomes stiffer when the embedded gap is closed (figure 2d). The surrounding arms are responsible for the soft mechanical response, and the cylindrical parts in the middle act as a mechanical fuse to protect the arms from further deformation and provide more resistance, resulting in higher stiffness and slower deformation. The 30 degrees inclinations of surrounding arms were considered for more flexibility compared to the case with straight arms while observing the 3D printing limit on the angle of overhangs with respect to the base plate (see section 5 for a detailed discussion).

In order to link the mechanical properties to a set of quantifiable parameters, the geometry of such a design consisting of one hexagonal cell (inside the box in figure 2a) should be uniquely expressed by a few parameters or measures. Except "D/2", which is essentially the bone rod radius constrained by the fractured bone size, they are manifested by ten parameters of a 2D sketch illustrated in figure 2a. The radii of curved edges (R1, R2, R3) provide the model with smooth stress flow to minimize stress concentration in those areas. The number of variables was chosen to allow for design flexibility to achieve different mechanical responses given the building material properties. Since the material should remain elastic due to cyclic loadings imposed on the bone rod, a linear curve was expected before the gap closure. The contact incident would result in a much stiffer response compared with the mechanical response before the gap closure. As a result, the desired force-displacement curve is bilinear, the first segment with a low slope (low stiffness) and the second one as close as possible to a vertical line (very high stiffness). This change is shown in figure 3 curve with additional visualization of FE results (contour plots of Mises stress in a 3D model).

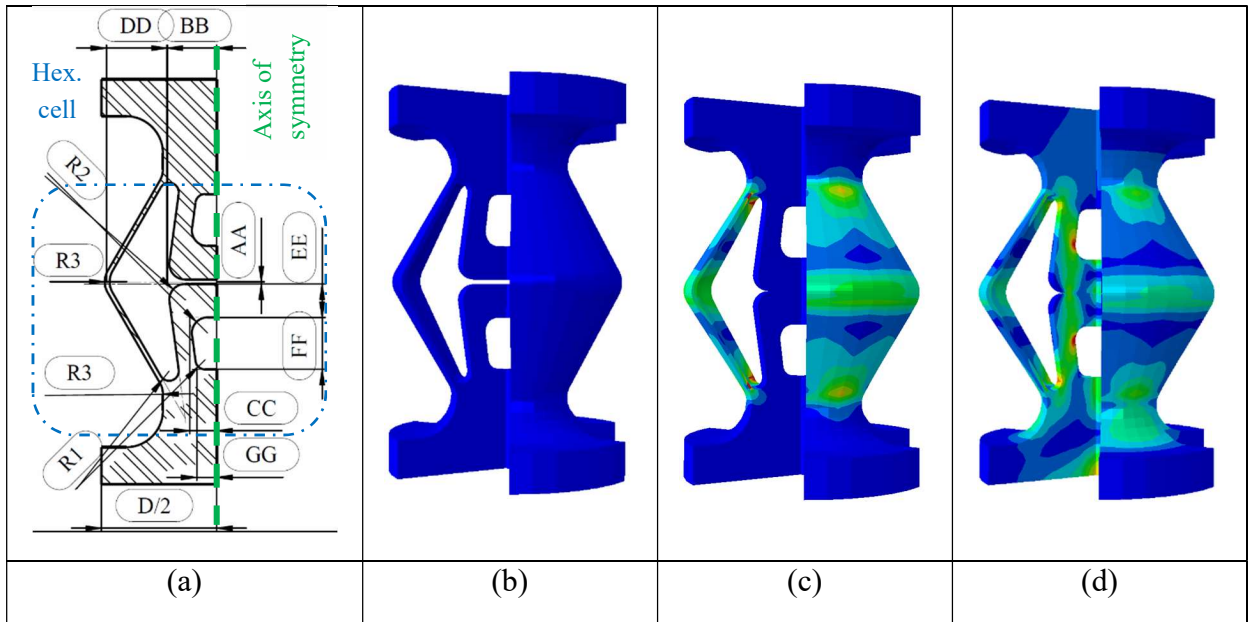

**Figure 2.** Conceptual design of a tunable stiffness metamaterial: a) 2D schematic of an axisymmetric design; b) Mises stress contour of a 3D model under no load, c) under axial load right before closing the gap, and d) under axial load after closing the gap.

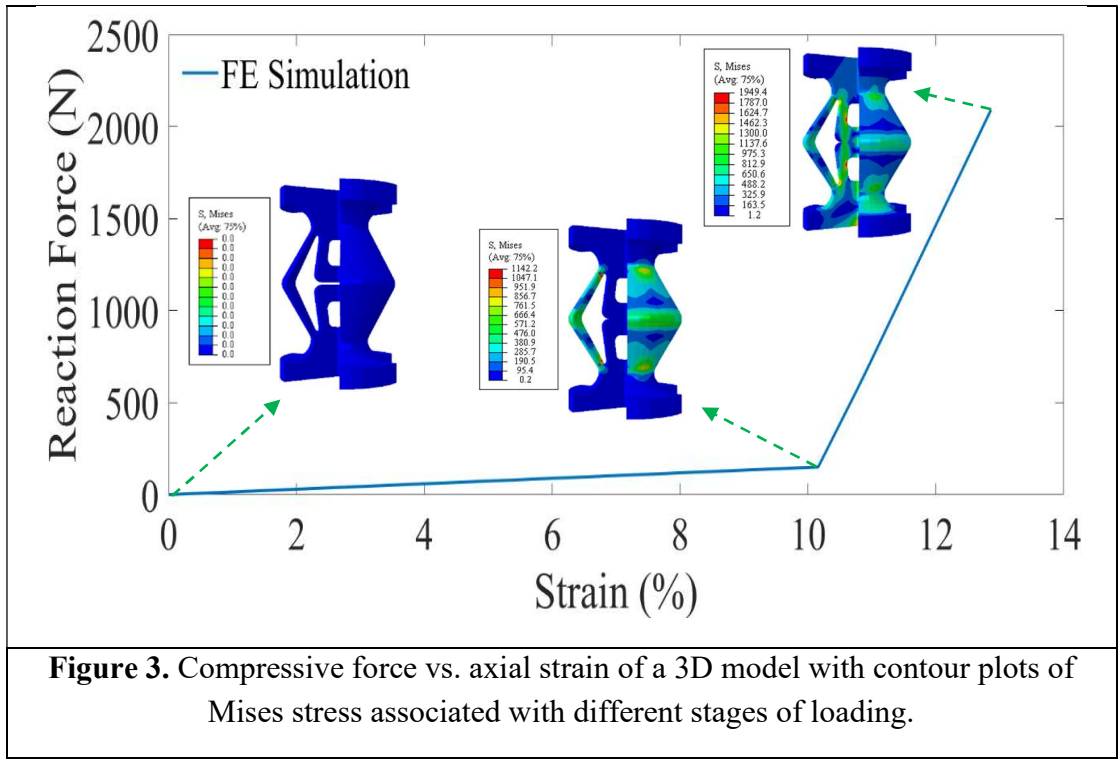

**Figure 3.** Compressive force vs. axial strain of a 3D model with contour plots of Mises stress associated with different stages of loading.

## 3. 3D Design of tunable stiffness bone rod

It is challenging to achieve a sufficiently low stiffness before the gap closure for common metals with high stiffness while maintaining the material yield limit. First, an axisymmetric model by a total revolution of the 2D sketch around the vertical symmetry axis was considered, as shown in the sketch figure (figure 2a). However, after several optimization trials, it was concluded that the part with just one hexagonal cell could not reach the curve points with a sufficiently low level of maximum Mises stress. Therefore, other designs were optimized with more hexagonal cells. Having more cells requires lower displacement in one cell since the total displacement is the sum of the displacements in all cells. As a result, the gap inside each cell should be narrowed accordingly. This may cause manufacturing problems due to limited accuracy in 3D printing or cutting methods if the cells are printed without the gap.

A partially revolved design, as shown in figure 4, can be made by 3D printing while leaving access areas for cutting tools. Thus, the manufacturer can continuously 3D-print the part without the gap, then remove the material from the gap or other sections. Another advantage of this design is the thicker walls, which is particularly desirable for manufacturing equipment of lower qualities, in surrounding arms.

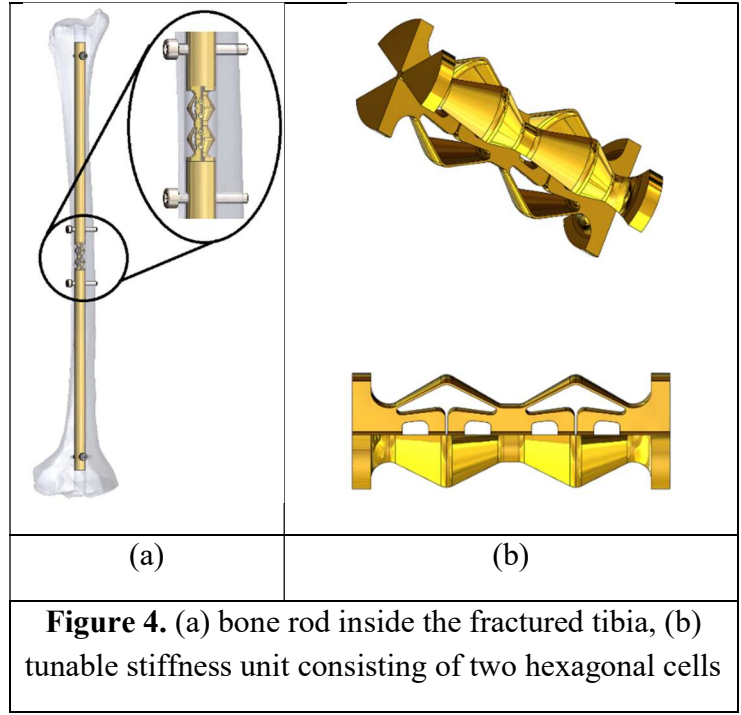

|  (a)  |  (b)  |
| --- | --- |

**Figure 4.** (a) bone rod inside the fractured tibia, (b) tunable stiffness unit consisting of two hexagonal cells

## 4. Design optimization
### 4.1. Defining objective functions

Different objective functions were proposed and tried for optimization. The initial attempts at finding the best designs considering only one objective function did not yield satisfying results. Therefore, two objective functions were used to optimize the design parameters, the first one having to do with curve characteristics, force-displacement behavior, and the second one having to do with elasticity measures.

One design parameter, the half gap size "AA" was fixed according to the fracture gap length to make sure that the rod becomes stiffer after an absolute relative displacement of the fractured part of the bone. As mentioned earlier, the optimum normal strain for bone growth by the osteoblast cells is about 9%. For instance, if the patient's fracture gap is 3 mm, the optimum compression in the fracture region should be about 0.27 mm. If the designed bone rod has two hexagonal cells aligned vertically, the total gap size inside each hex cell should then be equal to 0.135 mm since both are normally operating in the elastic region of the material before their gap closures. The remaining nine dimensional parameters, BB, CC, DD, EE, FF, GG, R1, R2, and R3, were the design variables in the optimization process.

The ratio of the maximum slope of the force-displacement curve to its respective minimum was considered as the first objective function, as shown in Eq. 1. However, the negative value of the ratio was computed to utilize optimization algorithms looking for minima of objective functions readily. Twenty equally spaced time intervals were considered in the FE process to obtain a simulated response with sufficient accuracy. This form of objective function dictates that design schemes showing more significant changes in their elastic behavior after the gap closure are more favorable in the optimization process.

$$\vec{d} = (AA = \alpha * fracture\_gap, BB, CC, DD, EE, FF, GG, R1, R2, R3)$$

$$OF_1(\vec{d}) = -\max_{1 \leq i \leq N_t} \{Slope_{Force-Disp}(\vec{d}, t(i)) > 0\} / \min_{1 \leq i \leq N_t} \{Slope_{Force-Disp}(\vec{d}, t(i)) > 0\} \quad (1)$$

In Eq. (1), $\vec{d}$ is the vector of design variables, and $\alpha * fracture\_gap$ is the value of parameter AA discussed earlier. $OF_1$ is the first objective function of design variables ($\vec{d}$), $N_t$ is the number of time steps of FE results, which contain the value of field outputs such as displacements and reaction forces, $Slope_{Force-Disp}$ is the slope of the force-displacement curve of a specific design ($\vec{d}$) determined at a specific time step of FE solution ($t(i)$), which is proportional to the total displacement imposed on the model.

The second objective was avoiding any plastic deformation, which would change the mechanical response permanently. Satisfying such a condition was first implemented by considering only the first objective and assigning an extreme value to this objective whenever the response of the structure surpasses the maximum allowable stress, which is the yield stress of the material. Single-objective optimization did not produce favorable results regardless of the optimization method used and its hyperparameters because better compliance with the objective curve points demands higher maximum Mises stress, especially in the surrounding arms with a thin layer of the material. Thus, a multi-objective optimization method was chosen to satisfy all design criteria simultaneously. Therefore, the maximum Mises stress in the whole design domain was considered as the second objective because the metallic materials used for the implants are mostly ductile. Only elastic behavior of the material was defined for the material properties in the FE simulation consistent with having stress levels lower than the yield limit of the material. Accordingly, the maximum Mises stress for every design scheme was extracted at the maximum possible force, proportional to the patient's weight, imposed on the fractured bone. It is formulated in Eq. 2 with $OF_2$ as the second objective function of the optimization problem, $N_{el}$ as the number of FE model elements, $\sigma_{Mises}$ as the Mises stress, $t$ as the time step, $RF(t)$ as the reaction force of the model due to compression, and $\beta * Weight$ as a fixed multiplier of patient's weight. $\beta$ was considered 1.3 as an average value according to previous studies, which assume increasing loading levels with the time of bone healing [5,10].

$$OF_2(\vec{d}) = \max_{1 \leq i \leq N_{el}} \{\sigma_{Mises}(\vec{d}, t, element_i) | RF(t) \leq \beta * Weight\} \quad (2)$$

FE modeling was performed to quantify the mechanical behavior of each design scheme in the optimization loop and compare it with the desired response of the structure. The FE analyses were done in the commercial software package of ABAQUS using mm, N, and tonne/m3 units for dimensions, forces, and densities, respectively. Simple elastic behavior was considered for material properties. Hard contact behavior was also defined in the gap area to simulate the contact behavior. All boundary conditions were defined as displacements for faster convergence. Therefore, the reaction force of the model in the direction of displacement was the output of the simulation. In the reversed case of force (pressure) loading, the output was model displacement.

### 4.2. Optimization algorithm and a case study

As the smooth behavior of the aforementioned objective functions has not been guaranteed, and finding the global extrema was desired in this study, the stochastic method of the GA for multi-objective functions was chosen for the optimization, a variant of NSGA-II as a controlled, elitist GA. The flowchart of the algorithm is shown in figure 5. The algorithm tries to find a non-dominated population of the design schemes. The Pareto front would illustrate such a population in the design variable space or the objective one as requested in the software. The objective variable space was more suitable to track the optimization process since it was 2D. The best point is the one showing the minimum values for the negative slope ratio of the force-displacement curve as well as maximum Mises stress in the FE simulation. However, these two have a reverse relationship, as shown in the Pareto fronts of optimization trials (see an example of the front in figure 6). The essential stopping criteria of NSGA-II is the spread quantity as a measure of Pareto set movement. Iterations will stop if the spread value change falls below a specific threshold while being less than an average of spreads in previous generations [22,39]. Other stopping criteria are the maximum number of iterations as well as a pre-determined set of threshold objectives. The threshold values were defined so that the optimization algorithm could stop searching for better design schemes if the first and second objectives were less than -10 as the negative slope ratio and the yield stress of the material, respectively. As a case study, the input data for modeling and design optimization were defined as presented in table 1 for a 70 kg patient with $\beta = 1.3$, and the last Pareto front was subsequently obtained, as shown in figure 6. The applied load on the model is higher than $\beta * Weight$ since stiffness change may occur in different load levels. "$r$" is the radius of a cylinder connecting the three revolved sections of the bone rod as shown in figure 4b. The best point of the front, for which the design parameters are presented in table 2, was also determined based on the constraint on the maximum allowable Mises stress (objective 2).

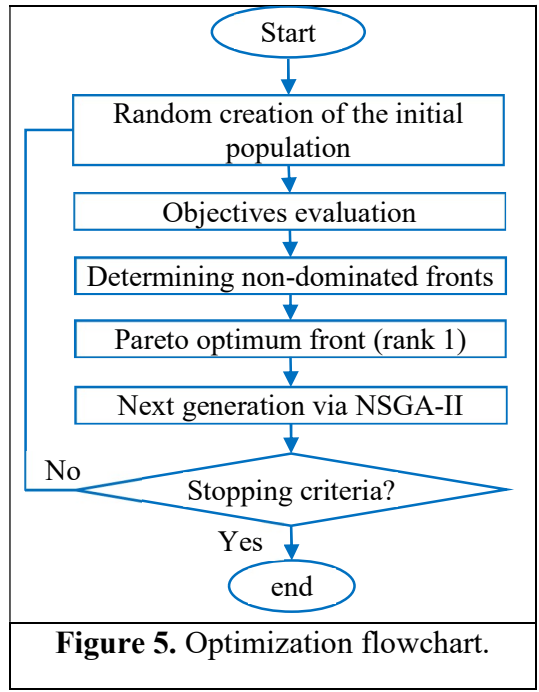

**Figure 5.** Optimization flowchart.

**Table 1.** Input data of optimization.

| D/2 (mm) | r (mm) | AA (mm) | Applied load (N) | Stiffness (GPa) | Poisson's ratio | Yield Stress (MPa) |
|---|---|---|---|---|---|---|
| 4 | 0.25 | 0.075 | 2100 | 200 | 0.27 | 1250 |

**Table 2.** Dimensional parameters of the best point in the Preto front (unit: mm).

| BB | CC | DD | EE | FF | GG | R1 | R2 | R3 |
|---|---|---|---|---|---|---|---|---|
| 1.615 | 0.883 | 1.975 | 1.084 | 1.691 | 0.655 | 0.314 | 0.507 | 0.715 |

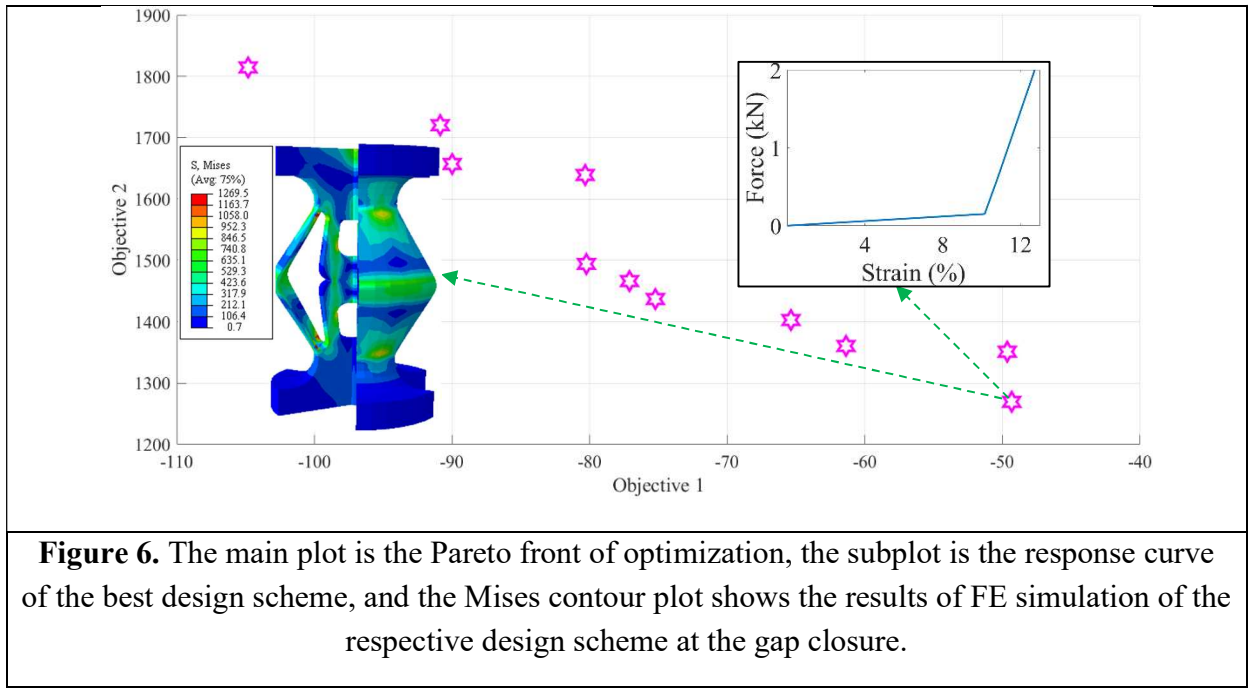

**Figure 6.** The main plot is the Pareto front of optimization, the subplot is the response curve of the best design scheme, and the Mises contour plot shows the results of FE simulation of the respective design scheme at the gap closure.

## 5. AM of a tunable stiffness bone rod

Powder bed fusion (PBF) is a metal AM process category [40] utilizing a thermal energy source that selectively sinters (i.e., without melting to the point of liquefaction) or fuses regions of a thin powder layer. Selective laser sintering as the first PBF method was later developed into selective laser melting and electron beam melting, which entirely melt small regions of metal powder called "melt pool" resulting in better mechanical properties than sintering method, due to technological and power source improvement. A thin layer of metal powder (typically between 30-50 μm) deposited on a flat substrate via a powder deposition system is locally melted along a predefined path to make the first layer of the part. Then, the substrate is lowered for a new powder layer to be deposited and transformed into the next slice of the part. This process will continue to reach the final slice [41].

The metal 3D printer used in this study was a 3D Systems ProX300, which utilizes fine steel powder (Stainless Steel 17-4 PH acquired from 3D Systems) and a 500-watt laser to melt the powder into the required geometry. It has a 250 mm * 250 mm build area, sufficient for 3D printing of the hexagonal cells of a complete bone rod. Having a compacting carbide roller, it increases material density and allows for larger overhangs, completely unsupported material, where the geometry allows [42]. It is especially useful when the cost of the whole manufacturing process and the effect of support material on the quality of the end product and on the complexity of removing them are taken into account. For example, the surrounding arms of our designed bone rod could be printed with minimum support material at about 60 degrees inclination with respect to the build plate illustrated in figure 7a with red sections as support material. However, the printing trials with the vertical orientation resulted in failed samples, as shown in the right side of the top part of figure 7b. The support material of the best horizontally printed part was machined off, while the surface roughness remained intact since it was not crucial for the application of this part inside the bone. The final machined part is shown in figure 7c. The digital measurement of gap size was 0.0053 in (0.1354 mm), which was lower than that of the original CAD model prepared for 3D printing (0.15 mm). The limited accuracy of 3D printing, as well as lower measurement accuracy due to surface roughness, are suspected to be the cause of this discrepancy.

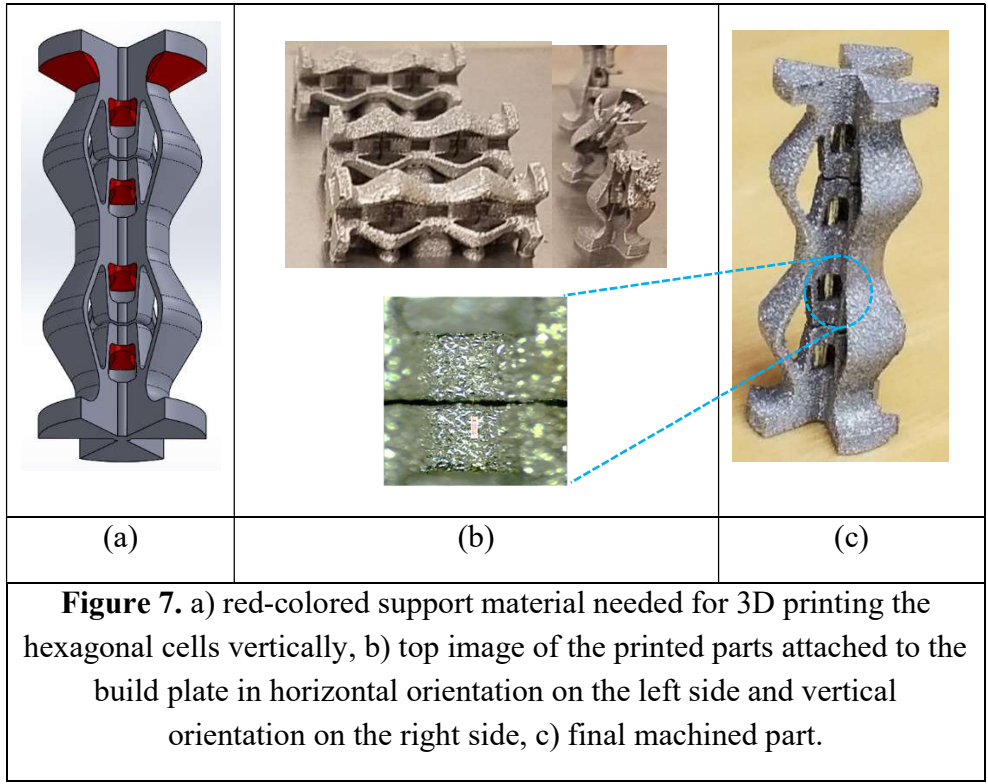

|(a)|(b)|(c)|

**Figure 7.** a) red-colored support material needed for 3D printing the hexagonal cells vertically, b) top image of the printed parts attached to the build plate in horizontal orientation on the left side and vertical orientation on the right side, c) final machined part.

The compression test was carried out using an Instron machine with a 20 kN load cell, and ASTM E9-19 standard method for compression testing of metallic materials at room temperature. The effective stiffness ratio was about 4.2, while FE simulation resulted in a higher slope ratio of 5. The possible reasons of the difference are the discrepancy between the printed material properties, elastic modulus and Poisson's ratio, and those defined in the FE simulation, limited accuracy of 3D printer in making a part with the exact dimensions of the CAD model, and plasticity due to surpassing the yield limit of material according to maximum Mises stress of the part in FE results. Therefore, our computational framework for designing the tunable stiffness bone rods will be efficient and accurate if the exact material properties are provided, and high loadings leading to material plasticity are avoided.

# 6. Conclusions

A new tunable stiffness bone rod was designed for healing tibia bone fracture. Upon applying axial force, bone rod deforms with low stiffness, and it becomes stiffer when the embedded gap gets closed. Considering the AM limitations, the patient-specific designs were obtained using multi-objective GA optimization in conjunction with FE modeling for calculating the objective functions, ensuring reversible and controlled elastic deformation. Such novel bone rods allow the broken bone to move in a controlled fashion along the longitudinal axis, and this motion stimulates bone healing while it prevents the common stress-shielding of ordinary bone rods, which leads to osteoporosis. Taking advantage of the flexible AM technique, the designed bone rods were 3D printed with the FDA approved material, stainless steel. The developed design framework and AM together offer optimum bone rods given patient weight, fracture gap, and selected building material. The main limitation of the practical use of this study lies in the AM end. A robust and accurate metal 3D printing should be utilized to transfer the optimum designs into real products with the intended optimum performance. Materials with lower stiffness and higher yield strength are much recommended since they lead to higher minimum thicknesses in surrounding arms of design while observing the elastic limit of the material for repeatable response in cyclic loading of the bone rod. Using finer powder in metal 3D printing also improves geometrical accuracy with a higher printing resolution.

**Acknowledgment** Authors thank Chris Hill and Jake J. Behrens from the Center for Industrial Research and Service, CIRAS, for 3D printing the parts, and sharing their expertise in AM.